\documentclass[preprint,showpacs,preprintnumbers,amsmath,amssymb]{revtex4}

\usepackage{graphicx}
\usepackage{dcolumn}
\usepackage{bm}

\begin{document}

\title{S-mixing and quantum tunneling of the magnetization in molecular nanomagnets}

\author{S. Carretta, E. Liviotti, N. Magnani, P. Santini and G. Amoretti}
\affiliation{Istituto Nazionale per la Fisica della Materia,
Dipartimento di Fisica, Universit\`a di Parma, I-43100 Parma,
Italy}

\date{January 21, 2004}
\begin{abstract}
The role of $S$-mixing in the quantum tunneling of the
magnetization in nanomagnets has been investigated. We show that
the effect on the tunneling frequency is huge and that the
discrepancy (more than 3 orders of magnitude in the tunneling
frequency) between spectroscopic and relaxation measurements in
Fe$_8$ can be resolved if $S$-mixing is taken into account.
\end{abstract}


\maketitle

Molecular nanomagnets (MNMs) \cite{Sessoli93,Gatteschi94} are
molecules containing transition-metal ions whose spins are so
strongly exchange-coupled that at low temperature each molecule
behaves like a single-domain particle with fixed total spin. One
of the most interesting phenomena displayed by these systems is
quantum tunneling of the {\it direction} of the total spin through
energy barriers\cite{Friedman96,Thomas96,Wernsdorfer99}. The
measured step-like magnetization curves of Mn$_{12}$ and Fe$_8$
provided macroscopic evidence of relaxation through quantum
tunneling. The latter is revealed by resonances observed in the
relaxation rate at specific values of the external magnetic field
$B_{AC}$, at which energy levels on opposite sides of the
anisotropy barrier are nearly degenerate and anticrossings (ACs)
in the field dependence of the energies occur. The relaxation rate
depends crucially on the value of the so-called tunnel splitting
$\Delta$, i.e. the gap at $B_{AC}$ between the quasi-degenerate
states. In particular, at very low temperature $T$ and at short
times $t$ the magnetization relaxes as $1-\sqrt{\Gamma t}$ where
$\Gamma \propto \Delta^2$ \cite{prokstamp}. For Fe$_8$, $\Delta$
was extracted in Ref. \cite{Wernsdorfer99} by measuring with a
microsquid apparatus the magnetization steps induced by sweeping a
longitudinal (i.e. parallel to the easy axis) applied field $B_z$
across $B_{AC}$. The size of these steps was linked to the tunnel
splitting through the Landau-Zener formula
$$
\Pi =1-e^{-\Delta^2/A},\eqno(1)
$$
where $\Pi$ is the tunneling probability at a level anticrossing,
and $A$ is proportional to the field sweeping rate. When the
experiment is performed in a static transverse field $B_y$,
$\Delta$ is found to display oscillations as a function of $B_y$,
which in a semiclassical approach reflect the destructive
interference of tunneling pathways.

Eq. (1) had been deduced by neglecting decoherence sources such as
hyperfine and dipolar fields. Nevertheless, it remains valid if
the sweeping rate is as fast as that actually used in the
experiments\cite{sinitsyn}. In addition, the model proposed in
Ref.\cite{loss} shows that the incoherent Zener tunneling can be
described by Eq. (1) with $\Delta$ renormalized by a factor
$\sqrt{2}$.

A striking circumstance is that the measured value of $\Delta$
seems completely incompatible with the value calculated by using
the Hamiltonian determined by inelastic neutron scattering
(INS)\cite{caciuffo98,amoretti00}, optical
spectroscopy\cite{mukhin} and electron paramagnetic
resonance\cite{hill02}. Indeed, the measured zero-field gap
$\Delta(B_y = 0)$ between the two lowest levels is near $10^{-7}$
K\cite{Wernsdorfer99}, while the value calculated ($4.44\cdot
10^{-11}$ K) is more than {\it 3 orders of magnitude} smaller.
This huge discrepancy seriously hinders any attempt to reach a
satisfactory theoretical modelling of the quantum tunneling of the
magnetization, where the square of $\Delta$ plays a fundamental
role.
The purpose of this work is to show that commonly neglected
quantum fluctuations of the {\it magnitude} of the total spin of
the molecule ({\it S-mixing} \cite{Liviotti02,epjb}) affect the
tunnel splitting of Fe$_8$ hugely, and allow the above-mentioned
discrepancy to be solved. Since Fe$_8$ displays a relatively small
degree of $S$-mixing, we expect the tunnel splitting of many
nanomagnets to be influenced even more heavily than in Fe$_8$ by
such fluctuations.

Each Fe$_8$ molecule can be described by the following spin
Hamiltonian:
$$
H=\sum_{i>j}J_{ij}{\bf s}_{i}\cdot {\bf s}_{j} +
\sum_{i}\sum_{k,q}b_k^q(i)O_k^q({\bf s_i}) +
$$
\vspace{-0.5cm}
$$
\sum_{i>j}{\bf s}_{i}\cdot {\bf D}_{ij}\cdot {\bf s}_{j}- g\mu_B
\sum_{i}{\bf B}\cdot {\bf s}_{i},\eqno(2)
$$
where ${\bf s}_{i}$ are spin operators of the $i^{th}$ Fe$^{3+}$
ion in the molecule ($s_i = 5/2$). The first term is the isotropic
Heisenberg exchange interaction. The second term describes the
local crystal-fields (CFs), with $O_k^q({\bf s_i})$ Stevens
operator equivalents for the $i$-th ion\cite{abragam} and
$b_k^q(i)$ CF parameters. Here $k=2$ or $4$ (larger values of the
rank $k$ are forbidden for $d$-electrons\cite{abragam}), and
$q=-k,...,k$. The third term represents the dipolar anisotropic
intra-cluster spin-spin interactions. The last term is the Zeeman
coupling with an external field ${\bf B}$. The exchange constants
$J_{ij}$ used in this work are those determined from
susceptibility\cite{barra2000}.

While the Heisenberg term is rotationally invariant and therefore
conserves the length $\vert {\bf S}\vert$ of the total spin ${\bf
S} = \sum_{i}{\bf s}_{i}$, the anisotropic terms do not conserve
this observable. Nevertheless, since the Heisenberg contribution
is usually largely dominant, $\vert {\bf S}\vert$ is nearly
conserved, and the energy spectrum of $H$ consists of a series of
level multiplets with an almost definite value of $\vert {\bf
S}\vert$. By neglecting the mixing between different
$S$-multiplets (i.e. $S$-mixing), the Hamiltonian Eq. (2) can be
projected onto each $S$-multiplet (strong-exchange limit):
$$
H_{sub}=B_2^0O_2^0{\bf (S)}+B_2^2O_2^2{\bf (S)}+B_4^0O_4^0{\bf
(S)}
$$
$$
+B_4^2O_4^2{\bf (S) }+B_4^4O_4^4{\bf (S)}- g\mu_B {\bf B}\cdot
{\bf S},\eqno(3)
$$
where ${\bf S}$ is a vector spin operator with $S$ equal to the
total spin of the $S$-multiplet\cite{nota}. The parameters $B_K^Q$
are calculated from $b_k^q(i)$ and ${\bf D}_{ij}$ by CF and
dipolar projection coefficients. This approach, applied to the
$S=10$ ground manifold of Fe$_8$, has allowed to interpret  INS
data very satisfactorily by assuming $B_2^0 = -9.75\times
10^{-2}K$, $B_2^2 = -4.66\times 10^{-2}K$,$B_4^0 = 1.0\times
10^{-6} K$,$B_4^2 = 1.2\times 10^{-7} K$,$B_4^4 = 8.6\times
10^{-6} K $ \cite{caciuffo98,amoretti00}. Very similar parameter
values are obtained from optical spectroscopy\cite{mukhin} and
electron paramagnetic resonance\cite{hill02}. In particular
$B_4^4$, which has the greatest effect on $\Delta$, is the same.
In order to reproduce the measured magnitude and oscillations of
$\Delta$ with Eq. (3), values of $B_K^Q$ incompatible with neutron
results had to be assumed\cite{Wernsdorfer99,loss,rastelli}. In
particular, $B_4^4$ was one order of magnitude larger and its sign
was reversed. In Fig. 1 we show calculations at $T= 9.6 K$ of the
INS spectrum of a Fe$_8$ powder with the experimental resolution
of Ref.\cite{caciuffo98}. The parameters used in
Refs.\cite{Wernsdorfer99,loss,rastelli} do not reproduce the INS
spectra satisfactorily, neither in the higher-energy part measured
in Ref.\cite{caciuffo98} nor in the lower-energy part measured
with the high resolution experiment of Ref.\cite{amoretti00}. If
the  strong-exchange-limit Hamiltonian (3) is used, there is no
way to reproduce with a unique set of parameters the magnitude and
behavior of $\Delta$ and the spectroscopic results. In the
following we show how this discrepancy can be removed if
$S$-mixing is taken into account. Indeed, although in Fe$_8$
$S$-mixing is a small perturbation (e.g., it produces negligible
changes in calculated spectroscopic quantities), its effect on
$\Delta$ is very large because it provides much more efficient
tunneling channels.

In order to evaluate $S$-mixing effects, we followed the method
developed in \cite{Liviotti02}, in which $S$-mixing is included up
to the second order in the anisotropy by a unitary transformation
applied to the complete Hamiltonian (2). The system can be still
described as an effective spin $S=10$, provided the
spin-Hamiltonian (3) is properly modified :  the parameters of the
Stevens operators are renormalized, and new higher rank ($K>4$)
terms are added. These latter are forbidden for $d$-electrons in
the strong-exchange limit. The advantage of using this method with
respect to large-scale numerical diagonalization (e.g. using the
Lanczos algorithm) is twofold: first of all, it allows calculation
times to be reduced drastically. In fact, the time-consuming part
of calculation (i.e. computing $\Xi$ and $\Upsilon$ in Eq. (9))
does not depend on the specific set of local CF parameters, and
therefore has to be performed only once. The second advantage is
that the simple and physically transparent single-spin formalism
of the strong-exchange limit is recovered.

Using as basis vectors the eigenvectors $\vert \alpha S M \rangle$
of the isotropic exchange $H_0$, the full Hamiltonian matrix $H$
Eq. (2) can be written as the sum of three terms
$$
H = H_0 + H_1 + H_2 \eqno(4)
$$
where $H_1 + H_2$ represents the anisotropic interactions. $H_1$
has nonzero elements only within the $S$ multiplets, while $H_2$
joins states with different $\alpha S$ and is the term responsible
for the mixing. $H_2$ is neglected in the strong-exchange limit
Eq. (3).

The perturbational procedure \cite{Liviotti02,Slicther64} consists
in performing a unitary transformation on $H$ such that the
off-diagonal (in $\alpha S$) blocks of the transformed Hamiltonian
$H^{\prime}$ are zero up to second order in the anisotropy. Hence,
in the new basis, states belonging to different multiplets are
uncoupled and the system can be described as an effective spin
multiplet, like in the strong-exchange limit. The matrix elements
of $H^{\prime}$ inside the ground multiplet $S=10$ are given by
$$
\langle S  M \vert H^{\prime} \vert S M^{\prime} \rangle =
E_{0}\delta_{M,M^{\prime}} + \langle \alpha S M \vert H_1 \vert
\alpha S M^{\prime} \rangle
$$
\vspace{-0.5cm}
$$
- \sum_{\alpha ^{\prime \prime} S^{\prime \prime} M^{\prime
\prime}} \frac{\langle \alpha S M \vert H_2 \vert \alpha ^{\prime
\prime} S^{\prime \prime} M^{\prime \prime} \rangle \langle \alpha
^{\prime \prime} S^{\prime \prime} M^{\prime \prime} \vert H_2
\vert \alpha S M^{\prime} \rangle }{E_{0 \alpha ^{\prime \prime}
S^{\prime \prime}} - E_{0}} \eqno(5)
$$
where $E_{0}$ is the lowest eigenvalue of $H_0$ and $\vert \alpha
S M \rangle$ are the corresponding eigenvectors. $\vert \alpha
^{\prime \prime} S^{\prime \prime} M^{\prime \prime}\rangle$ are
excited eigenvectors of $H_0$ with energy $E_{0 \alpha ^{\prime
\prime} S^{\prime \prime}}$.

The second term in Eq. (5) coincides with the strong-exchange
Hamiltonian (3), while the last term represents mixing
corrections. By exploiting the Wigner-Eckart
theorem\cite{Liviotti02} the latter can be written in general as
$$
\sum_{K,Q} {\tilde B}_K^Q O_K^Q \eqno(6)
$$
with $K\le 8$ and even, and $-K\le Q\le K$. Hence
$$
H_1+H_2=\sum_{K,Q} C_K^Q O_K^Q \eqno(7)
$$
where $C_K^Q=B_K^Q +{\tilde B}_K^Q$ ($B_K^Q=0$ for $K>4$).
Therefore, on the one hand $S$-mixing introduces in the effective
Hamiltonian new terms forbidden in the strong-exchange limit, and
on the other hand it renormalizes the coefficients of the other
terms (with $K\le 4$). This implies that the CF parameters
determined by INS are to be regarded as $C_K^Q$s rather than as
the $B_K^Q$s of Eq. (3).

The parameters ${\tilde B}_K^Q$ are given by linear combinations
of products of reduced matrix elements. For example,
$$
{\tilde B}_6^6=\sum_{i,j} b_2^2(i) b_4^4(j)\Xi(i,j) + \sum_{i}
b_4^4(i)\Upsilon(i) \eqno(8)
$$
with \vspace{-0.5cm}
$$
\Xi(i,j)=\sum_{S^{\prime \prime}=S-2}^{S+2}c_{66}^{S-S^{\prime
\prime}}\cdot
$$
\vspace{-0.5cm}
$$
\sum_{\alpha ^{\prime \prime}} \frac{\langle \alpha S \vert \vert
T ^{(2)}(i)\vert \vert \alpha ^{\prime \prime} S^{\prime \prime}
\rangle \langle  \alpha S\vert \vert T^{(4)}(j)\vert \vert \alpha
^{\prime \prime} S^{\prime \prime}\rangle} {E_{0 \alpha ^{\prime
\prime} S^{\prime \prime}} - E_{0}} \eqno(9a)
$$
and \vspace{-0.5cm}
$$
\Upsilon(i)= -\sqrt{2}\sum_{j,k}J^u_{jk}\sum_{S^{\prime
\prime}=S-2}^{S+2}c_{66}^{S-S^{\prime \prime}}\cdot
$$
\vspace{-0.5cm}
$$
 \sum_{\alpha ^{\prime \prime}} \frac{\langle
\alpha S \vert \vert T ^{(2)}( 11\vert jk)\vert \vert \alpha
^{\prime \prime} S^{\prime \prime} \rangle \langle  \alpha S\vert
\vert T^{(4)}(i)\vert \vert \alpha ^{\prime \prime} S^{\prime
\prime}\rangle} {E_{0 \alpha ^{\prime \prime} S^{\prime \prime}} -
E_{0}}. \eqno(9b)
$$
$i$, $j$ and $k$  label magnetic ions, $T^{(K)}(i)$ and $T ^{(2)}(
11\vert jk)$ are the tensor operators describing the local CF and
dipole-dipole interactions\cite{spagnoli99}. $J^u_{jk}$ are
defined in terms of the elements of ${\bf D}_{ij}$ (Eq. (2)) in
Ref. \cite{spagnoli99}. The $c_{66}^{S-S^{\prime \prime}}$
coefficients are defined according to the theory developed
in\cite{Liviotti02}. Expressions similar to (8), (9a) and (9b) are
obtained for the other parameters ${\tilde B}_K^Q$, which are all
expressed as polynomials of $2^{nd}$-order in the $b_k^q(i)$.
Hence $S$-mixing gives rise to {\it highly efficient} and
otherwise forbidden tunneling channels, by generating new
high-rank anisotropy terms.

In order to assess the impact of these terms on the
tunnel-splitting we applied our theory quantitatively. While the
dipole-dipole interaction (appearing, e.g., in (9b) through
$J^u_{jk}$ ) can be computed by the point-dipole
approximation\cite{Bencini90}, the local CF parameters $b_k^q(i)$
cannot be determined {\it ab initio} reliably. Therefore, by
numerically inverting the $2^{nd}$-order functions
$C_K^Q(\{b_k^q(i)\})$, we determine the possible sets
$\{b_k^q(i)\}_f$ consistent with INS, i.e. such that the values
$C_K^Q(\{b_k^q(i)\}_f)$ for $K\le 4$ coincide with those
determined by INS (within experimental error bars). Even by
neglecting all $b_k^q(i)$ with $q\ne 0,2,4$ (i.e. those not
contributing to $H_1$) and by enforcing on the $\{b_k^q(i)\}$ the
approximate $D_2$ molecular symmetry of Fe$_8$
\cite{notasimmetria}, there are still more unknown parameters than
constraints, and we find therefore that there are infinitely many
sets compatible with INS. For these sets, the distribution of the
calculated values of the tunnel splitting $\Delta$ is shown in
Fig. 2a by a histogram of the log-increments $\log
(\Delta/\Delta_0)$. Here the local CF parameters $b_k^q(i)$ vary
on a grid bounded by $\vert b_2^q(i)\vert < 8$ K, $\vert
b_4^q(i)\vert < 0.4$ K. This choice is based on two
considerations: {\it i)} the experimental values of $C_2^0$ and
$C_2^2$ set only a lower bound $\vert b_2^q(i)\vert \agt 1 K$.
Local second-order parameters of the order of few $K$ are
reasonable in case of $Fe^{3+}$ in a low symmetry environment
\cite{muller}. {\it ii)} Typical ratios of fourth- to second-order
CF parameters range (in modulus) between 0.01 and 0.1. Fig. 2a
shows that $S$-mixing plays a crucial role since typically it
enhances $\Delta$ by several orders of magnitude. Fig. 2b reports
the result of the same calculation when the grid bounds are
restricted to $\vert b_2^q(i)\vert < 4$ K, $\vert b_4^q(i)\vert <
0.04$ K. Even with this more restrictive choice the effect of
$S$-mixing remains huge. The measured value of $\Delta$ is
indicated by an arrow and falls well inside the distribution.

We have shown that $S$-mixing can remove the discrepancy between
the value of the zero-field gap $\Delta (B_z=0,B_y =0)$ measured
by relaxation experiments and that calculated from the
spectroscopic results. Now it remains to prove that also the
measured oscillations of $\Delta (B_z=0,B_y)$, and the behavior of
the AC gaps $\Delta_{ex}(B_z=B_{AC},B_y)$ between excited states
are reproducible. The aim of this work is not to perform a
best-fit of the observed oscillations of the tunnel-splittings,
but to prove that $S$-mixing eliminates the inconsistency between
spectroscopic and Landau-Zener measurements. Therefore, we limit
Eq.(6) to values of $K\le 6$ in order to find among the infinite
possible parameter sets consistent with INS, one involving as few
high-rank terms as possible, and reproducing the AC gaps behavior
satisfactorily . The new terms in Eq.(7) (forbidden in the
strong-exchange limit) with significative influence on $\Delta$
are then $C_6^4$ and $C_6^6$.

As a first step, we fixed $C_K^Q$ for $K\le 4$ to the values
determined by neutron spectroscopy (reported below Eq. (3)) and we
chose values of $C_6^4$ and $C_6^6$ which reproduce the behavior
of $\Delta$ satisfactorily. With $C_6^4\sim -1.8\times 10^{-7}$ K
and $C_6^6\sim -1.15\times 10^{-7}$ K, $\Delta$ calculated at zero
applied field is $\sim 1.1\times 10^{-7}$ K, to be compared with
the value $\sim 0.4\times 10^{-10}$ K obtained when $C_6^4=0$ and
$C_6^6=0$. Moreover, the measured oscillations of $\Delta$ as a
function of the transverse field $B_y$ are well reproduced, as
well as the AC gaps involving the excited states $\vert
-10\rangle$ and $\vert 10-n\rangle$ ($n=1$,$2$) (see Fig. 3). The
nonzero value of experimental oscillations at minima may arise
from experimental artifacts, (e.g., crystal mosaicity). The
behavior of $\Delta$ as a function of the transverse field modulus
for nonzero azimuthal angles $\phi$ between the applied field and
the $y$-axis, is also in good agreement with
measurements\cite{Wernsdorfer99}.

As a second step, we checked that the addition of these 6th-order
terms does not affect the INS cross-section significantly. In
fact, the recalculated cross-section is indistinguishable from
that calculated in \cite{caciuffo98} and reported in Fig.1, apart
from an irrelevant shift (by $\sim 20$ $\mu$eV) of the shoulder at
0.16 meV.

Values of $C_6^4$ and $C_6^6$ of the order and sign of those given
above are realistic in Fe$_8$. For example, the insets in Figs. 2a
and 2b show the distribution $P(C_6^6)$ of values of $C_6^6$
calculated on the same grids as described above. We stress that
our choice of high-rank parameters is merely the simplest
possibility. There are many different sets involving also the
other high-rank terms which would be consistent with experimental
data. Although a unique determination of the $C_K^Q$s is not
possible, the important point is that the addition of high-rank
terms, which is allowed only if $S$-mixing is considered, is
essential to describe consistently relaxation and spectroscopic
data.

In conclusion, we have shown that the discrepancy (more than 3
orders of magnitude in the tunnel splitting) between spectroscopy
and relaxation measurements in Fe$_8$ can be resolved if
$S$-mixing is taken into account. Even a small degree of
$S$-mixing has huge influence in the tunneling dynamics since it
opens highly efficient tunnel channels through otherwise forbidden
high-rank anisotropy terms. The degree of $S$-mixing is strongly
influenced by the topology of the molecule. Therefore in addition
to the height of the anisotropy barrier, also the cluster topology
must be taken into account in designing new nanomagnets.

\newpage

FIGURE CAPTIONS:

Fig.1: Calculated INS intensity for a Fe$_8$ powder with the
spin Hamiltonian Eq. (3) and various parameter sets. The energy
resolution has been fixed to the experimental value of
19$\mu$eV\cite{caciuffo98}. The parameters used in
Ref.\cite{rastelli} are close to those used in
Ref.\cite{Wernsdorfer99}and yield almost the same intensity curve.
The inset shows a schematic view of Fe$_8$.

Fig.2: Calculated distribution of values of the tunnel splitting
$\Delta (B_z=0,B_y=0)$ normalized to the value $\Delta_0 =
4.44\cdot 10^{-11}$ K obtained without $S$-mixing (with the
parameters obtained from INS). The local CF parameters $b_k^q(i)$
vary on grids (different in (a) and (b)) defined in the text.
Arrows indicate the measured ratios. Insets show the distribution
of values of $C_6^6$ on the same grids.

Fig.3: Top: measured tunnel splitting as function of an applied
transverse magnetic field $B_y$ with $B_z =0$ ($n=0$), and AC gaps
involving the excited states $\vert -10\rangle$ and $\vert
10-n\rangle$ ($n=1$,$2$) with $B_z = B_{AC} \sim n\cdot 0.22$ T
($n=1,2$) \cite{Wernsdorfer99}. Bottom: the same quantities
calculated with the Hamiltonian (7) and the $C_K^Q$ parameters
given in the text.


\begin{references}
\bibitem{nota} Other second- and fourth-order terms are expected
to be negligibly small because of the approximate $D_2$ symmetry
of the molecule.
\bibitem{notasimmetria} $b_k^q(1)=b_k^q(2)$, $b_k^q(3)=b_k^q(4)$,
$b_k^q(5)=b_k^q(6)=b_k^q(7)=b_k^q(8)$, $q=0,2,4$ (indices are
defined in the inset of Fig. 1). To further reduce the number of
free parameters we assumed $b_4^4(1)=b_4^4(3)$.
\end{references}
\end{document}